\documentclass[12pt]{amsart}
\usepackage{graphicx}
\usepackage{amsfonts}
\usepackage{amsmath}
\usepackage{cases}
\usepackage{amsfonts}
\usepackage{amssymb}
\usepackage[margin=1in]{geometry}
\usepackage{mathrsfs}
\usepackage{verbatim}
\usepackage{hyperref}
\usepackage{datetime}
\usepackage{amstext}
\usepackage{amsxtra}
\usepackage{amsopn}
\usepackage{amsthm}

\theoremstyle{definition}

\theoremstyle{plain}

\theoremstyle{remark}

\theoremstyle{plain}

\theoremstyle{plain}

\theoremstyle{plain}

\renewcommand{\P}{\mathbb{P}}

\begin{document}

\title[Conformal field theories and compact curves in moduli spaces]{Conformal field theories and \\compact curves in moduli spaces}
\author{Ron Donagi}
\address{Department of Mathematics\\David Rittenhouse Lab.\\
University of Pennsylvania\\
209 South 33rd Street\\Philadelphia, PA  19104-6395}
\email{donagi@math.upenn.edu}

\author{David R. Morrison}
\address{Departments of Mathematics and Physics\\
University of California, Santa Barbara\\
Santa Barbara, CA 93106}
\email{drm@math.ucsb.edu}

\date{Revised March 20, 2018}

\begin{abstract}
We show that there are many compact subsets of the moduli space $M_g$ of Riemann surfaces of genus $g$ that do not intersect any symmetry locus.  This has interesting implications for $\mathcal{N}=2$ supersymmetric conformal field
theories in four dimensions.
\end{abstract}

\maketitle

\section{Introduction}

An interesting analysis of conformal manifolds for conformal
field theories in even spacetime dimension
was presented in \cite{KS}, using properties of the trace anomaly.
 For supersymmetric theories whose conformal
manifold is K\"ahler, including the cases of $\mathcal{N}=(2,2)$
supersymmetry in two dimensions\footnote{In the $\mathcal{N}=(2,2)$ case,
the argument applies  only to one ``half'' of the conformal manifold
at a time:  either
the half whose tangent space consists of chiral marginal operators,
or the half whose tangent space consists of twisted chiral marginal
operators.}  and $\mathcal{N}=2$ supersymmetry
in four dimensions,  it was predicted in \cite{KS} that the 
Zamolodchikov metric on the conformal manifold (i.e., the metric
determined by the two-point
functions of the conformal theory) should have a globally-defined
K\"ahler potential, closely related to the partition function on the
sphere as conjectured in \cite{JKLMR}.  
The mathematical implications of this prediction
are quite remarkable.  In particular, in this case
the conformal manifolds in question
could not contain compact holomorphic curves.

In two dimensions, a familiar class of examples of $\mathcal{N}=(2,2)$
theories is given by sigma models whose target is a Calabi--Yau manifold $X$
of fixed topological type, and one ``half'' of the conformal manifold
is represented by the moduli space $\mathcal{M}_X$ of complex structures on the
Calabi--Yau manifold.  The prediction that the K\"ahler potential on this
conformal manifold is
globally well-defined is equivalent to claiming that the Hodge line bundle on $\mathcal{M}_X$, whose fiber at a given $X$ is the one-dimensional space of holomorphic 3-forms on $X$, admits constant, $U(1)$-valued transition functions. For several families of Calabi--Yau manifolds, this prediction has been verified in a recent paper \cite{DS}.
Other aspects of this prediction are analyzed in \cite{MP}.

In four dimensions,
the $SU(2)$ theories of class $S$ with no punctures \cite{G,GMN},
given by compactifying the $(2,0)$ theory in six dimensions of type $A_1$
on a Riemann surface of some fixed genus $g$,
form a collection of $\mathcal{N}=2$ supersymmetric conformal field
theories whose conformal manifold precisely corresponds to
the moduli space ${M}_g$ of Riemann surfaces
of genus $g$.
This $3g-3$ dimensional space is non compact. 
The prediction from \cite{KS} that the Zamolodchikov metric%
\ on $M_g$ has a globally
well-defined K\"ahler potential implies that $M_g$ contains no compact
analytic subspaces.  As we shall show, this is false.  In fact,
no K\"ahler metric on $M_g$ can have a globally well-defined
K\"ahler potential.

In conversations with some of the authors of \cite{KS}, it was realized
that at points of the conformal manifold at which the Riemann surface
(and hence the conformal field theory) has extra symmetries, the
extra symmetries might interfere with the original prediction.
One can instead
consider the open subset $M_g^0 \subset M_g$ parametrizing those Riemann surfaces without extra symmetries, i.e., those that do not admit any holomorphic automorphisms. The  prediction arising from \cite{KS} would then be that this space $M_g^0$ contains no compact analytic subspaces of positive dimension. In this note we show that, on the contrary, $M_g^0$ for $g>3$ does contain lots of compact subspaces. In fact, through every point of $M_g^0$, or even any finite collection of its points, there is such a compact subspace. 

There were two developments
after the first version of this paper was circulated.  
First, Joerg Teschner pointed out to us that 
the generalization of \cite{Pestun:2007rz,Alday:2009aq} 
from linear quivers to
general UV curves proposed in \cite{Vartanov:2013ima} identifies
the K\"ahler potential
of the Zamolodchikov metric\footnote{That is, the four-sphere partition
function \cite{KS}.} with the Liouville
partition function on Riemann surfaces of genus $g$.
The Liouville partition function is not globally well-defined
on $M_g$, but is rather the K\"ahler potential
of a hermitian metric on a non-trivial projective line
bundle over $M_g$ when $g>1$, in much the same way as first argued by
Friedan and Shenker for rational CFTs \cite{Friedan:1986ua}.\footnote{In 
addition,
a very direct relation between the classical Liouville partition function and
the Weil--Petersson K\"ahler form follows from  results of Takhtajan
and Zograf \cite{MR889594}.
Yuji Tachikawa has pointed out to us that this should be regarded as
a type of ``large $N$ limit'' of Teschner's observation.}
The result of the present paper shows that this K\"ahler potential
cannot be well-defined even over the subset $M_g^0$ of $M_g$ when $g>3$.

Second, motivated by the 
example which this paper contains, Tachikawa and Yonekura 
\cite{Tachikawa:2017aux}
showed that in four dimensions,
the trace anomaly mentioned above contains  previously-overlooked 
terms of the form
$c_2(R)
\,
[\omega^Z]/2\pi$
and
$p_1\,[\omega^Z]/2\pi$, where $c_2(R)$ is the second Chern class
of the background $SU(2)_R$ gauge field, $p_1$ is the first
Pontryagin class of the spacetime metric,  and $[\omega^Z]$ is the
cohomology class of the Zamolodchikov metric on the conformal
manifold.
The existence of these terms resolves the apparent conflict between
\cite{KS} and the result we present below.  Further details will
appear in \cite{STY}.

Our precise claim is that there are (lots of) compact curves in $M_g$, for $g >2$, 
and that most of those miss the loci of enhanced symmetry, if $g>3.$
The gist of the argument is as follows. 
Even though we are interested in $M_G$, or rather its open subset $M_g^0$, 
it will be convenient to work temporarily with a compactification.
We construct a $3g-3$-dimensional projective variety $\overline{M_g}^{S}$ 
that contains $M_g$ as a dense open subset.
In it, we need to control two types of loci: the boundary
$\overline{M_g}^S \setminus M_g$, and 
the loci of Riemann surfaces with a symmetry.
The technical part of the argument shows that all of these loci have codimension $\geq 2$, 
so dimension $\leq 3g-5$. 
Since $\overline{M_g}^{S}$ is a projective variety, 
it can be embedded in projective spaces  $\P^N$ for arbitrarily large $N$.
Given a finite set $S$ of points of $M_g^0$, choose a large enough $N$
and $3g-4$ generic hypersurfaces containing the image of $S$ in $\P^N$. 
The locus $B$ cut out by these hypersurfaces will have codimension $3g-4$ in $\P^N$. 
If these hypersurfaces are sufficiently generic, 
this $B$ will meet the image of $\overline{M_g}^{S}$
in a 1-dimensional compact submanifold,
and will meet neither the boundary
$\overline{M_g}^S \setminus M_g$ nor 
the loci of Riemann surfaces with a symmetry.
This produces the desired compact curve in $M_g^0$ containing the specified set $S$.

\section{The codimension of the boundary}

The best known compactification of $M_g$ is
the Deligne-Mumford compactification 
$\overline{M_g}^{DM}$ of $M_g$, \cite{DM},
parametrizing stable curves.
Its boundary $\overline{M_g}^{DM} \setminus M_g$ 
is too big - each of its components has codimension 1. 
We need a different compactification 
in which the boundary is blown down to a smaller dimensional locus.
This is provided by the Satake-type compactification $\overline{M_g}^{S}$ that we describe below.

As is well known, the boundary 
$\overline{M_g}^{DM} \setminus M_g$ 
consists of smooth, normal crossing divisors $\Delta_i, \  i \geq 0$.
All the the $\Delta_i, i \geq 0$, are divisors, i.e. codimension 1 loci, 
in the Deligne-Mumford compactification $\overline{M_g}^{DM}$. 
For $i>0$, $\Delta_i$ parametrizes reducible curves $C=C' \cup C''$.
At a generic point of $\Delta_i$, 
$C'$ is a smooth curve of genus $i$ and $C''$ is a smooth curve of  genus $g-i$,
and they intersect transversally at one point, forming a node there.
At special points of $\Delta_i$, 
$C'$ and $C''$ can themselves be stable curves, 
still of genera $i$ and $g-i$ respectively.
At a generic point of $\Delta_0$, 
the curve $C$ is nodal but now irreducible. 
It can be obtained from a curve $C'$ of genus $g-1$ 
(called the normalization of $C$) 
by gluing two points of $C'$ to each other, transversally.
It is easy to count that the dimension of each $\Delta_i$ equals $3g-4$.

We let $A_g$ denote the moduli space of $g$-dimensional, principally polarized abelian varieties.
This is a quotient of Siegel's space, parametrizing {\em{period matrices}},
or symmetric $g \times g$ complex matrices with positive imaginary part,
by the integral symplectic group.
Its dimension is $g(g+1)/2$.
The moduli space $A_g$ has its own compactification,
the Satake compactification $\overline{A_g}^S$.
This is a compact complex space, 
which as a set can be identified with the disjoint union
$ \cup_{i=0}^{g} A_i$
of the $A_i$ for $i\leq g$.
Its boundary $\overline{A_g}^S \setminus A_g$ is small:
it is the disjoint union
$ \cup_{i=0}^{g-1} A_i$
of the $A_i$ for $i<g$.
It has codimension $g$ in $\overline{A_g}^S$.

The Abel-Jacobi map $AJ: M_g \to A_g$ 
sends a Riemann surface $C$ to its Jacobian variety $J(C)$,
which is the abelian variety specified by the period matrix of $C$. 
Torelli's theorem tells us that $AJ$ is injective,
so (ignoring some irreleavnt minor complication along the hyperelliptic locus),
we can identify $M_g$ with its image $AJ(M_g)$.
It turns out that the  Abel-Jacobi map extends to a map
${\overline{AJ}}: \overline{M_g}^{DM} \to \overline{A_g}^S$
from the Deligne-Mumford compactification $\overline{M_g}^{DM}$ 
to the Satakel compactification $\overline{A_g}^S$.
The Satake compactification $\overline{M_g}^{S}$ of $M_g$  is the image of ${\overline{AJ}}$. 
It is our alternative compactification of $M_g$.

This extended Abel-Jacobi map ${\overline{AJ}}$
is no longer an embedding:
it collapses the boundary divisors. 
For $i>1$, the divisor (i.e. codimension 1 locus) $\Delta_i$
collapses to a locus of codimension 3 in $\overline{M_g}^{S}$.
This locus is actually in the interior of $A_g^S$, not in its boundary:
The image points describe compact, i.e. non-singular, abelian varieties, 
that happen to be products of two lower dimensional abelian varieties,
namely the Jacobians of $C',C''$.
Equivalently, the period matrix of such an abelian variety is block diagonal, 
the two blocks being the periods of $C',C''$.
The fibers of ${\overline{AJ}}$ on these $\Delta_i$ are two dimensional: 
the locations of the two glued points are ignored by ${\overline{AJ}}$. 

For $i=1$, $\Delta_1$ still goes to the locus of products, 
in the interior of $A_g$,
but its image is larger:
each elliptic tail decreases the codimension by 1. 
When $g>2$ there can be only
one elliptic tail, so the codimension of the image in $\overline{M_g}^{S}$ is 2, 
and the fibers are one dimensional: 
the location of the non elliptic glued point is ignored.
When $g=2$ there are two elliptic tails, and the codimension of the image is 1.
This is why we need to assume $g>2$.

The divisor $\Delta_0$, on the other hand, 
goes to a locus in the boundary of $\overline{A_g}^S$:
A nodal curve $C$ with normalization $C'$ goes to the Jacobian of $C'$, 
which is in the boundary 
$\overline{A_g}^S \setminus A_g = \cup_{i=0}^{g-1} A_i$.
The image of $\Delta_0$ is therefore the locus of genus $g-1$ Jacobians;
its dimension is therefore $3g-6$,
so it has codimension 3 in the Satake compactification $\overline{M_g}^{S}$.

All in all, we see that for $g>2$, the Satake compactification $\overline{M_g}^{S}$
has a boundary of codimension $\geq 2$ as desired.

\section{The codimension of the symmetry locus}

Next, here is the count showing that the codimension in $M_g$ of the
symmetry loci is at least 2 whenever $g>3$.

Let $g$ be the genus of a (smooth) curve C with automorphism $a: C \to C$ of
order $d$. Let $g_0$ be the genus of the quotient curve $C_0 := C /<a>$, and let
$b$ be the number of branch points. The number of moduli of curves with such
automorphisms is $3g_0-3 + b$, because given $C_0$ and the branch locus there is
a finite number of such $C$'s.

The Hurwitz relation gives: $2g-2 \geq d(2g_0-2) + db/2$. So the codimension
in moduli of the locus of curves $C$ with enhanced symmetry is at least:
\[
3g-3 - (3g_0-3 + b)   \geq
(3/2) (d(2g_0-2) + db/2) - (3g_0-3 + b) =
(3d-3)(g_0-1) + (3d/4 - 1)b
\]

If $g_0>1$ this codimension is at least $3d-3 \geq 3$.

If $g_0=1$, this is $(3d/4 - 1)b$, which is at least 2
unless $d=2$ and $b=2$ so $g=2$.

If $g_0 =0$, our lower bound is 
\[
3bd/4 - b -3d + 3 =
(3/4)[(d-(4/3))(b-4)]-1. 
\]
This can be $\leq 1$ only if
$[(d-(4/3))(b-4)] \leq 8/3$. So:
\begin{itemize}
\item either $d=2$ and $b \leq 8$, so $g \leq 3$ (This is the case of
hyperelliptic curves, which indeed have codimension 1 in genus 3);
\item or $b \leq 5$, so the dimension of the enhanced symmetry locus is at
most $5-3=2$, so the codimension in $M_g$ is at least 4 unless $g \leq 2$.
\end{itemize}

We thank Richard Hain for pointing out that calculations similar to the content of the present section have appeared elsewhere, e.g.\ in Proposition 5.1 of 
\cite{MR2784328}.

\section{Conclusion}

We have seen that, as long as $g>3$,
all components of both the boundary and the symmetry loci  
have codimension at least 2 in $\overline{M_g}^{S}$,
so dimension at most $3g-5$.
As we saw in the introduction, 
after we embed $\overline{M_g}^{S}$ in a large projective space $\P^N$,
it follows that the intersection of $\overline{M_g}^{S}$ 
with $3g-4$ generic hypersurfaces in $\P^N$ 
is a smooth, compact 1-dimensonal space 
that avoids both the boundary and the loci of enhanced symmetry. 
Its inverse image in $M_g$ is then 
a smooth, compact curve, contained in $M_g$ and avoiding the enhanced symmetry loci.
Clearly we can choose our hypersurfaces to pass through 
any point or finite collection of points of $M_g^0$.

As a consequence of this mathematical analysis, we conclude that it
is not possible for any K\"ahler metric on $M_g^0$ (including the
Zamolodchikov metric) to have a globally-defined K\"ahler potential.
This provides additional evidence for the validity of the
 modification made in \cite{Tachikawa:2017aux,STY}
to the original analysis of \cite{KS}.

\section{Acknowledgements}
We are grateful to Jaume Gomis, Richard Hain,
Zohar Komargodski, Ronen Plesser,
Nati Seiberg, Eric Sharpe, Yuji Tachikawa, and Joerg Teschner
for correspondence and conversations.
RD was supported in part by 
NSF grant DMS-1603526 and by Simons Foundation grant \# 390287.
DRM was supported in part by NSF grant PHY-1620842.


\begin{thebibliography}{99}

\bibitem{KS}
 J. Gomis, P.-S. Hsin, Z. Komargodski, A. Schwimmer, N. Seiberg,
S. Theisen,
``Anomalies, conformal manifolds, and spheres,''
JHEP {\bf 1603} (2016) 022,
{\tt arXiv:1509.08511 [hep-th]}.

\bibitem{JKLMR}
  H.~Jockers, V.~Kumar, J.~M.~Lapan, D.~R.~Morrison and M.~Romo,
  ``Two-Sphere partition functions and Gromov-Witten invariants,''
  Commun.\ Math.\ Phys.\  {\bf 325} (2014) 1139,
  {\tt arXiv:1208.6244 [hep-th]}.


\bibitem{DS}
R. Donagi, E. Sharpe,
``On the global moduli of Calabi-Yau threefolds,''
{\tt arXiv:1707.05322 [math.AG]}.

\bibitem{MP} D. R. Morrison and M. R. Plesser, to appear.

\bibitem{G} 
  D.~Gaiotto,
  ``N=2 dualities,''
  JHEP {\bf 1208} (2012) 034,
  {\tt arXiv:0904.2715 [hep-th]}.

\bibitem{GMN}
  D.~Gaiotto, G.~W.~Moore and A.~Neitzke,
  ``Wall-crossing, Hitchin systems, and the WKB approximation,''
Adv.\ Math.\ {\bf 234} (2013) 239--403,
  {\tt arXiv:0907.3987 [hep-th]}.

\bibitem{Pestun:2007rz} 
  V.~Pestun,
  ``Localization of gauge theory on a four-sphere and supersymmetric Wilson loops,''
  Commun.\ Math.\ Phys.\  {\bf 313}, 71 (2012),
  {\tt arXiv:0712.2824 [hep-th]}.

\bibitem{Alday:2009aq} 
  L.~F.~Alday, D.~Gaiotto and Y.~Tachikawa,
  ``Liouville Correlation Functions from Four-dimensional Gauge Theories,''
  Lett.\ Math.\ Phys.\  {\bf 91}, 167 (2010),
  {\tt arXiv:0906.3219 [hep-th]}.

\bibitem{Vartanov:2013ima} 
  J.~Teschner and G.~S.~Vartanov,
  ``Supersymmetric gauge theories, quantization of $\mathcal{M}_{\mathrm{flat}}$, and conformal field theory,''
  Adv.\ Theor.\ Math.\ Phys.\  {\bf 19}, 1 (2015),
  {\tt arXiv:1302.3778 [hep-th]}.

\bibitem{Friedan:1986ua} 
  D.~Friedan and S.~H.~Shenker,
  ``The Analytic Geometry of Two-Dimensional Conformal Field Theory,''
  Nucl.\ Phys.\ B {\bf 281}, 509 (1987).

\bibitem{MR889594}
P.~G. Zograf and L.~A. Takhtadzhyan, ``On the uniformization of {R}iemann
  surfaces and on the {W}eil-{P}etersson metric on the {T}eichm\"uller and
  {S}chottky spaces,'' Mat. Sb. (N.S.) {\bf 132 (174)} (1987) 304--321, 444;
  English translation in Math. USSR-Sb. {\bf 60} (1988),  297--313.

\bibitem{Tachikawa:2017aux} 
  Y.~Tachikawa, K.~Yonekura,
  ``Anomalies involving the space of couplings and the Zamolodchikov metric,''
  {\tt arXiv:1710.03934v3 [hep-th]}.

\bibitem{STY} 
N. Seiberg, Y. Tachikawa, K. Yonekura,
        ``Anomalies of Duality Groups and Extended Conformal Manifolds,''
        to appear.

\bibitem{DM}
P. Deligne, D. Mumford, 
``The irreducibility of the space of curves of given genus,'' Publ.\ 
Math.\ IH\'ES {\bf 36} (1969) 75--109.

\bibitem{MR2784328}
R.~Hain, ``Rational points of universal curves,'' J. Amer. Math. Soc. {\bf
  24} (2011) 709--769, {\tt arXiv:1001.5008 [math.NT]}.




\end{thebibliography}
\end{document}